
\input phyzzx

\def\df{\varphi}

\def\bg{{\hat g}}
\def\bgr{{\rm e}^{2\rho} {\hat g}}
\def\bgs{{\rm e}^{2\sigma} {\hat g}}
\def\blap{{\hat \Delta}}
\def\bcv{{\hat R}}
\def\vr{\delta \rho}

\def\vg{\delta g}
\def\vp{\delta \df}
\def\vf{\delta f}
\def\vx{\delta \xi}
\def\dg{\hbox{$\sqrt{-g}$}}
\def\dbg{\hbox{$\sqrt{-{\hat g}}$}}
\def\e{{\rm e}}
\def\det{{\rm det}}

\def\pp{\prime}
\def\pd{\partial}

\REF\cghs{C. Callan, S. Giddings, J. Harvey and A. Strominger, Phys. Rev.
          {\bf D45} (1992) R1005.}
\REF\rst{J. Russo, L. Susskind and L. Thorlacius, Phys. Lett. {\bf 292B}
         (1992) 13;
         T. Banks, A. Dabholkar, M. Douglas and M. O'Loughlin, Phys. Rev
         {\bf D45} (1992) 3607.}
\REF\w{S. Hawking, Phys. Rev. Lett. {\bf 69} (1992) 406;
       L. Susskind and L. Thorlacius, Nucl. Phys. {\bf B382} (1992) 123;
       B. Birnir, S. Giddings, J. Harvey and A. Strominger, Phys. Rev.
       {\bf D46} (1992) 638.}
\REF\h{K. Hamada, {\it Quantum Theory of Dilaton Gravity in 1+1 Dimensions},
       \break preprint UT-Komaba 92-7.}
\REF\str{A. Strominger, {\it Fadeev-Popov Ghosts and 1+1 Dimensional Black
        Hole Evaporation}, preprint UCSB-TH-92-18;
        S. deAlwis, Phys. Lett. {\bf 289B} (1992) 278;
        A. Bilal and C. Callan, {\it Liouville Models of Black Hole
        Evaporation}, preprint PUPT-1320;
        S. Giddings and A. Strominger, {\it Quantum Theories of Dilaton
        Gravity}, preprint UCSB-TH-92-28.}
\REF\dk{J. Distler and H. Kawai, Nucl. Phys. {\bf B321} (1989) 509;
        F. David, Mod. Phys. Lett. {\bf A3} (1988) 1651.}
\REF\fr{D. Friedan, In {\it Recent Advances in Field Theory and Statistical
        Mechanics} ({\it Les Houches, 1982}), edited by J. Zuber and
        R. Stora, North-Holland.}
\REF\al{O. Alvarez, Nucl. Phys. {\bf B216} (1983) 125.}
\REF\ct{T. Curtright and C. Thorn, Phys. Rev. Lett. {\bf 48} (1982) 1309;
        N. Seiberg, Prog. Theor. Phys. Suppl. {\bf 102} (1990) 319.}
\REF\tih{P. Thomi, B. Isaak and P. Hajicek, Phys. Rev. {\bf D30} (1984) 1168;
         P. Hajicek, Phys. Rev. {\bf D30} (1984) 1178.}

\pubnum{UT-Komaba 92-9}

\titlepage

\title{{\bf Gravitational Collapse in 1+1 Dimensions  \break
              and Quantum Gravity}\footnote\dagger{Talk given at \lq\lq YITP
              Workshop on Theories of Quantum Fields -Beyond Perturbation-",
              Kyoto, Japan, 14-17 July 1992.}}

\author{Ken-ji Hamada}

\address{Institute of Physics, University of Tokyo \break
         Komaba, Meguro-ku, Tokyo 153, Japan \break}

\centerline{\bf Abstract}

     We investigate the quantum theory of 1+1 dimensional dilaton
gravity, which is an interesting toy model of the black hole dynamics. The
functional measures of gravity part are explicitly evaluated and derive the
Wheeler-DeWitt like equations as physical state conditions. In ADM formalism
the measures are very ambiguous, but in our formalism they are explicitly
defined. Then the new features which are not seen in ADM formalism come out.
A singularity appears at $\df^2 =\kappa (>0) $, where
$\kappa =(N-51/2)/12 $ and $ N$ is the number of matter fields. At the final
stage of the black hole evaporation, the Liouville term becomes important,
which just comes from the measures of the fields. Behind the
singularity the quantum mechanical region $\kappa > \df^2 >0 $ extends, where
the sign of the kinetic term in the Wheeler-DeWitt like equation changes.
If $\kappa <0 $, the singularity disappears. We briefly discuss the
possibility of gravitational tunneling and the issue of the information loss.

{\bf 1. Introduction}

   Recently many authors investigate the dynamics of the black hole by using
an interesting toy model of gravity in 1+1 dimensions.\refmark{\cghs-\str}
It is called the dilaton gravity, which was
proposed by Callan, Giddings, Harvey and Strominger.\refmark{\cghs}
The model has very
similar features to the spherically symmetric gravitational system in 3+1
dimensions. The essence of the black hole dynamics appears to be included
enough. Really in the semi-classical approximation we can argue the dynamics
in completely parallel way to the case of the spherically symmetric
black hole. Furthermore the gravitational back-reaction effects can be
included systematically.

   In this paper we develop the argument to the quantum gravity.\refmark{\h}
The quantum
gravity becomes very important at the final stage of the black hole
evaporation. It is expected that the issue of the information loss may be
resolved in the quantum gravity.

   As a quantization method of gravitation, there is  Arnowitt-Deser-Misner
(ADM) formalism or Wheeler-DeWitt approach. However there are some problems in
ADM formalism, the issues of measures and orderings. Here we explicitly
evaluate the contributions of measures. Following the procedure of
David-Distler-Kawai (DDK)\refmark{\dk}
we determine the measures of metrics in conformal
gauge. From the gauge fixed theory the physical state conditions are derived.
Then the new features which are not seen in  ADM formalism appear.

{\bf 2. Quantum dilaton gravity}

     The theory of 1+1 dimensional dilaton gravity is defined by the
following
action\footnote\dagger{Compare with the spherically symmetric
gravitational system in 3+1 dimensions. If the metric is restricted as
$ds^2 = g_{\alpha\beta}dx^{\alpha} dx^{\beta} + \df^2 d\Omega^2$, where
$\alpha, \beta =0, 1 $ and
$d\Omega^2 $ is the volume element of a unit 2-sphere, the Einstein-Hilbert
action becomes
$$
   I_{EH}  = {1 \over 16\pi} \int d^4 x \hbox{$\sqrt{-g^{(4)}}$} R^{(4)}
           = {1 \over 4} \int d^2 x \hbox{$\sqrt{-g}$}
                (R \df^2 +2 g^{\alpha\beta}
                         \pd_{\alpha} \df \pd_{\beta} \df +2) ~.
$$
The resemblance to the dilaton gravity is manifest. Thus the image of $\df=r $
is very convenient when we consider the dynamics of the dilaton gravity.}
$$
\eqalign{
       &  I(g,\df,f)=I_D (g,\df)+I_M (g,f) ~,               \cr
       &  I_D (g,\df)= {1 \over 2\pi} \int d^2 x \dg
               (R \df^2
                +4 g^{\alpha \beta} \partial_{\alpha} \df \partial_{\beta} \df
                +4 \lambda^2 \df^2 ) ~,        \cr
       &  I_M (g,f) = -{1 \over 4\pi} \sum^N_{j=1} \int d^2 x \dg
           g^{\alpha \beta} \partial_{\alpha} f_j \partial_{\beta} f_j ~, \cr
        } \eqno(1)
$$
where $\df ={\rm e}^{-\phi} $ is the dilaton field and $f_j $'s are $N$ matter
fields. $\lambda^2 $ is the cosmological constant.  $R $ is the
curvature of the metrics $g $. The classical equations of motion can be solved
exactly and one obtains, for instance, the black hole geometry
$$
        \df^2 = {\rm e}^{-2\rho} = {M \over \lambda} -\lambda^2 x^+ x^- ~,
          \qquad f_j =0 ,
        \eqno(2)
$$
where $g_{\alpha \beta}={\rm e}^{2\rho} \eta_{\alpha \beta} $,
$\eta_{\alpha \beta} =(-1,1) $ and $x^{\pm} =x^0 \pm x^1 $. $M $ is the mass
of the black hole. More interesting geometry is the gravitational collapse.
It is given by
$$
       \df^2 =\e^{-2\rho} =
         -{M \over \lambda x^+_0 } (x^+ -x^+_0 ) \theta (x^+ -x^+_0 )
            -\lambda^2 x^+ x^-  ~,
       \eqno(3)
$$
where the infalling matter flux is given by the shock wave along the line
$x^+ =x^+_0 $
$$
     {1 \over 2} \sum^N_{j=1} \pd_+ f_j \pd_+ f_j
        = {M \over \lambda x^+_0 } \delta (x^+ -x^+_0 )  ~.
    \eqno(4)
$$

    The quantum theory of the dilaton gravity is defined by
$$
    Z = \int {D_g(g) D_g(\df) D_g(f) \over {\rm Vol}(Diff.) }
           {\rm e}^{iI(g,\df,f)} ~,
     \eqno(5)
$$
where ${\rm Vol}(Diff.) $ is the gauge volume. The functional measures are
defined
from the following norms\footnote\star{The definitions of measures in
ref.5 are quite different from ours. Their definitions are mysterious for me,
especially the origin of the $b-c $ ghosts. Thus our quantum theory appears to
be
quite different from theirs.}
$$
\eqalign{
    &  < \vg, \vg >_g =
           \int d^2 x \dg g^{\alpha \gamma} g^{\beta \delta}
              \vg_{\alpha \beta} \vg_{\gamma \delta} ~,        \cr
    &  < \vp, \vp>_g =
           \int d^2 x \dg \vp \vp  ~,                           \cr
    &  < \vf_j , \vf_j >_g =
           \int d^2 x \dg \vf_j \vf_j
             \qquad (j=1, \cdots N)  ~.                       \cr
          }   \eqno(6)
$$

   Let us first discuss the measure of the metrics. We decompose the metrics
into a conformal factor $\rho $ and a background metric $\bg $ as
$ g = \bgr $. This is the conformal gauge-fixing condition adopted here.
The change in the
metric is given by the change in the conformal factor $\vr $
and the change under a
diffeomorphism $ \vx_{\alpha}$ as
$$
\eqalign{
          \vg_{\alpha \beta}
             & = 2\vr g_{\alpha \beta}
               + \nabla_{\alpha} \vx_{\beta}
               + \nabla_{\beta} \vx_{\alpha}           \cr
             & = 2 \vr^{\pp} g_{\alpha \beta}
               + (P_1 \vx )_{\alpha \beta} ~,          \cr
         }    \eqno(7)
$$
where
$$
       \vr^{\pp} = \vr + {1 \over 2} \nabla^{\gamma} \vx_{\gamma} ~,
         \qquad
       (P_1 \vx)_{\alpha \beta} = \nabla_{\alpha} \vx_{\beta}
                  + \nabla_{\beta} \vx_{\alpha}
                  - g_{\alpha \beta} \nabla^{\gamma} \vx_{\gamma} ~.
      \eqno(8)
$$
The variations $ \vr^{\pp} g_{\alpha\beta} $ and $ (P_1 \vx)_{\alpha\beta} $
are orthogonal in the functional space defined by the norms (6). Therefore
the measure over metrics can be decomposed as
$$
\eqalign{
         D_g (g) & = D_g (\rho^{\pp}) D_g (P_1 \xi)    \cr
                 & = D_g (\rho)  D_g (\xi_{\alpha}) \det_g P_1  ~.    \cr
        }   \eqno(9)
$$
The functional integration over $\xi_{\alpha} $ cancels out the gauge volume.
The Jacobian $\det_g P_1 $ can be represented by  the functional integral
over the ghosts $b, c$. Thus the partition function (5) becomes
$$
     Z = \int D_g (\rho) D_g (\df) D_g (f) D_g (b) D_g (c)
           \exp \bigl[ iI_D (g,\df)+iI_M (g,f)+iI_{gh}(g,b,c) \bigr]  ~,
      \eqno(10)
$$
where  $I_{gh} $ is the well-known ghost action (see for
example  ref.7). The measure $D_g (\rho) $ is defined from the norm (6) by
$$
    <\vr, \vr>_g = \int d^2 x \hbox{$\sqrt{-g} $} (\vr)^2
                 = \int d^2 x \hbox{$\sqrt{-\bg} $} \e^{2\rho} (\vr)^2 ~.
     \eqno(11)
$$

   This is not the end of the story.  The expression (10) has serious problems.
The measure (11) is not invariant under the local shift
$\rho \rightarrow \rho +h $ and also the measures of the fields $\df, f, b$
and $c $ explicitly depend on the dynamical variable $g =\bgr $. This is quite
inconvenient because we must pick up contributions from the measures when
the conformal factor $\rho $ is integrated. So we will rewrite  the measures
on $ g $ into more convenient ones  defined on the background metric $\bg $.

    We do not repeat the calculation in detail, which was discussed in
ref.4.
Here we mention  the outline of the arguments and list  some key relations.
First we rewrite the measures
of the dilaton, the matter and the ghost fields into the convenient ones.
It is realized by using the well-known transformation property for the
measures of the matter and the ghost fields (see for example ref.8)
$$
     D_{\bgr}(f) D_{\bgr}(b) D_{\bgr}(c)
          = \exp \biggl[ i{N-26 \over 12\pi} S_L (\rho, \bg) \biggr]
            D_{\bg}(f) D_{\bg}(b) D_{\bg}(c)
     \eqno(12)
$$
and the relation for the measure of the dilaton field
$$
        \int D_{\bgr} (\df) \e^{iI_D (\bgr,\df)}
           = \exp \biggl[ i{c_{\df} \over 12\pi} S_L (\rho, \bg) \biggr]
               \int D_{\bg}(\df) \e^{iI_D (\bgr, \df)}
        \eqno(13)
$$
with $c_{\df} = -1/2 $, which was proved in ref.4. $S_L (\rho, \bg) $ is
what is called the Liouville action defined by
$$
        S_L (\rho, \bg) = {1 \over 2}
                  \int d^2 x \dbg ( \bg^{\alpha \beta}
                   \partial_{\alpha} \rho \partial_{\beta} \rho
                   + \bcv \rho ) ~.
          \eqno(14)
$$
(Note that the actions of the matter and the ghost fields are invariant
under the Weyl rescalings or $ I_M (g, f) =I_M (\bg, f) $ and
$I_{gh}(g,b,c)=I_{gh}(\bg,b,c) $, but the action of the dilaton part is
not so. Pay attention to the $\rho $-dependence of each side of (13).)
{}From eqs. (12) and (13) we get
$$
\eqalign{
   Z =  \int D_{\bgr}(\rho) & D_{\bg}(\df) D_{\bg}(f) D_{\bg}(b) D_{\bg}(c)
          \exp \biggl[ i {c_{\df}+N-26 \over 12\pi} S_L (\rho, \bg)      \cr
          &  + iI_D (\e^{2\rho} \bg,\df) +iI_M (\bg,f)
               +iI_{gh}(\bg,b,c) \biggr] ~.                          \cr
        }   \eqno(15)
$$

    Next we rewrite the measure of $\rho $. According to the procedure of
DDK,\refmark{\dk} we assume the following relation
$$
       D_{\bgr} (\rho) = D_{\bg} (\rho)
            \exp \biggl[ i {A \over 12\pi} S_L (\rho, \bg) \biggr] ~.
       \eqno(16)
$$
Note that the measure $D_{\bg} (\rho) $ is invariant under the local shift of
$\rho $. The parameter $A $ is determined by the consistency. Since the
original theory depends only on the metrics $g= \bgr $, the theory should be
invariant under the simultaneous shift
$$
        \rho \rightarrow \rho - \sigma ~, \qquad \bg \rightarrow \bgs ~.
          \eqno(17)
$$
This requirement leads to $A=1 $.  Finally we get the expression
$$
      Z = \int D_{\bg} (\Phi) \e^{i{\hat I}(\bg,\Phi)} ~,
         \eqno(18)
$$
where $\Phi $ denotes the fields $\rho, \df, f, b $ and $c $. ${\hat I} $ is
the gauge-fixed action
$$
\eqalign{
   {\hat I} = & {1 \over 2\pi} \int d^2 x \dbg \Bigl[
    4{\hat g}^{\alpha \beta} \partial_{\alpha} \df \partial_{\beta} \df
    +4{\hat g}^{\alpha \beta} \df \partial_{\alpha} \df \partial_{\beta} \rho
       + \bcv \df^2 +4 \lambda^2 \df^2 \e^{2\rho}          \cr
      & + \kappa (
         {\hat g}^{\alpha \beta} \partial_{\alpha} \rho \partial_{\beta} \rho
         + \bcv \rho )
       -{1 \over 2} \sum^N_{j=1}
         {\hat g}^{\alpha \beta} \partial_{\alpha} f_j \partial_{\beta} f_j
       \Bigr] +I_{gh}(\bg,b,c)                             \cr
        }  \eqno(19)
$$
with
$$
      \kappa ={1 \over 12}(1+c_{\df}+N-26)={N-51/2 \over 12} ~.
        \eqno(20)
$$

    Closing this section there are some remarks. We showed that the theory
(which includes the
measures) is invariant under the simultaneous shift (17). Furthermore  the
measure $D_{\bg} (\rho)$ is invariant under the local shift of $\rho $. So the
theory is invariant under conformal changes of the background metric $\bg $:
$\bg \rightarrow \bgs$. More explicitly the Liouville-dilaton part is
transformed as
$$
\eqalign{
        & \int D_{\bgs} (\rho) D_{\bgs}(\df)
              \exp \biggl[ i{\kappa \over \pi} S_L (\rho, \bgs)
                      +iI_D (\e^{2\rho}\bgs,\df) \biggr]        \cr
        & = \int D_{\bgs} (\rho) D_{\bgs}(\df)
              \exp \biggl[ i{\kappa \over \pi} S_L (\rho-\sigma, \bgs)
                      +iI_D (\bgr,\df) \biggr]        \cr
        & = \exp \biggl[ i{1+c_{\df} \over 12\pi}S_L(\sigma,\bg) \biggr]
            \int D_{\bg} (\rho) D_{\bg}(\df)
              \exp \biggl[ i{\kappa \over \pi} S_L (\rho-\sigma, \bgs)
                      +iI_D (\bgr,\df) \biggr]        \cr
        & = \exp \biggl[ -i{N-26 \over 12\pi}S_L(\sigma,\bg) \biggr]
            \int D_{\bg} (\rho) D_{\bg}(\df)
              \exp \biggl[ i{\kappa \over \pi} S_L (\rho, \bg)
                      +iI_D (\bgr,\df) \biggr]  ~,      \cr
       }   \eqno(21)
$$
where in the last equality we use the relation for the Liouville action
$$
       S_L (\rho-\sigma, \bgs) = S_L (\rho,\bg)-S_L (\sigma,\bg) ~.
        \eqno(22)
$$
The extra Liouville action $-i{N-26 \over 12\pi} S_L(\sigma,\bg)$ cancels
out with that induced from the measures of the matter and ghost fields
(see eq.(12)) so that the partition function is invariant under the conformal
change of $\bg $.
The exact proof is given in ref.4. This invariance is quite reasonable
because the background metric $\bg $ is very artificial. The theory should be
independent of how to choose the background metric.

    Here there is a question whether the theory (18) is regarded as a kind
of conformal field theory (CFT) on $\bg $ or not.  Usual definition of CFT is
that the action is invariant under the conformal transformation. According to
this definition the Liouville theory is not CFT. However, as shown in ref.9,
the energy-momentum tensors of the quantum Liouville theory satisfy the
Virasoro algebra.
So it is considered as a kind of CFT. In the theory (18), if we
ignored the coupling between the Liouville field $\rho$ and the dilaton field
$\df $, the Liouville part would be regarded
as CFT with the central charge $c_{\rho} = 1-12\kappa $. In general CFT is
described by a set of decoupled fields, while the theory (18) has the
non-trivial coupling so that it is quite different from usual
CFT.\footnote\dagger{Furthermore note that CFT generally indicates a theory
which is conformally invarint at the classical level, but not at the quantum
level by anomalies. On the other hand, in the case of the dilaton gravity,
it is meaningless to discuss the invariance under the conformal change of
$\bg $ at the classical level.
It is significant only in the quantum gravity.}

   The second remark is that the partition function is a scalar.
This is manifest in the definition (5). After rewriting the partition
function into the expression (18), however, this invariance is hidden.
It is instructive to
show that the partition function is really scalar. The Liouville field $\rho $
is transformed as
$$
      \rho^{\pp} (x^{\pp})=\rho (x) -\gamma (x) ~, \qquad
        \gamma (x) = {1 \over 2} \log
                     \biggl\vert {\pd x^{\pp} \over \pd x} \biggr\vert^2  ~,
      \eqno(23)
$$
where we only consider the conformal coordinate transformation
$x^{\pm \pp} =x^{\pm \pp}(x^{\pm}) $ to preserve the conformal gauge and use
the notation $\vert x \vert^2 = x^+ x^- $. On the other hand
the background metric is not transformed: $\bg^{\pp}(x^{\pp}) =\bg (x) $.
It is natural because the background metric is not dynamical.
Therefore the gauge-fixed action is transformed as
$ {\hat I}^{\pp} ={\hat I}-{\kappa \over \pi} S_L (\gamma, \bg) $, where note
that$R $ is a scalar, but $\bcv $ is transformed as
$\bcv^{\pp} =\vert {\pd x \over \pd x^{\pp}}
\vert^2 (\bcv +2\blap \gamma) $. The measures defined on $\bg $ are also
non-invariant under the coordinate transformation. The extra Liouville term
$S_L (\gamma,\bg) $ cancels out with that coming from the measures so that
the partition function is invariant.\footnote\star{Note that after all the
invariance under the conformal change of $\bg $ is in other words the
invariance under the coordinate transformation.}

{\bf 3. Physical state conditions}

     Now we carry out the canonical quantization of the gauge-fixed 1+1
dimensional dilaton gravity. As mentioned in Sect.2 the theory
should be independent of how to choose the background metric $\bg $. Thus
the variation of the partition function with respect to $\bg $ vanishes
$$
    0= {\delta Z \over \delta {\hat g}^{\alpha \beta} }
     = \int D_{{\hat g}} (\Phi)
            {\delta {\hat I} \over \delta {\hat g}^{\alpha \beta} }
            {\rm e}^{i{\hat I}({\hat g},\Phi)}
      + \int {\delta D_{{\hat g}} (\Phi) \over \delta {\hat g}^{\alpha \beta}}
            {\rm e}^{i{\hat I}({\hat g},\Phi)} ~.
    \eqno(24)
$$
The first term of r.h.s. is nothing but
$< {\delta {\hat I} \over \delta {\hat g}^{\alpha \beta} } >_{{\hat g}} $.
The second term picks up an anomalous contribution.
But if we choose the Minkowski background ${\hat g} =\eta $,
this contribution vanishes. So it is convenient to choose the Minkowski
background metric. Thus
the physical state conditions are
$$
       \langle {\delta {\hat I} \over \delta \bg^{\alpha \beta}}
            \rangle_{\bg =\eta} =0
           \eqno(25)
$$
or
$$
      < {\hat T}_{00} >_{\bg=\eta} = <{\hat T}_{01} >_{\bg=\eta} =0 ~,
            \eqno(26)
$$
where the energy-momentum tensor ${\hat T}_{\alpha \beta} $ is defined by
${\hat T}_{\alpha \beta}=-{2 \over \dbg}{\delta {\hat I} \over \delta
\bg^{\alpha\beta}} \vert_{\bg=\eta}$.
The condition for ${\hat T}_{11} $ reduces to the
one for $ {\hat T}_{00} $ by using the $\rho $-equation of motion.
Furthermore we restrict the physical state to the one which satisfies the
condition $<{\hat T}^{gh}_{\alpha\beta}>_{\bg=\eta}=0$ because the ghost flux
should vanish in the flat space time.

   Since the functional measures are defined on the Minkowski background
metric, we can set up the canonical commutation relations in usual way.
The conjugate momentums for $\df,\rho $ and $f_j $ are given by
$$
\eqalign{
     \Pi_{\df} &= -{4 \over \pi} {\dot \df}
                       -{2 \over \pi} \df {\dot \rho}  ~,      \cr
     \Pi_{\rho}&= -{\kappa \over \pi} {\dot \rho}
                       -{2 \over \pi} \df {\dot \df} ~,        \cr
     \Pi_{f_j} &= {1 \over 2\pi}  {\dot f}_j  ~,               \cr
        } \eqno(27)
$$
where the dot stands for the derivative with respect to the
time coordinate. Then the physical state conditions (26) can be expressed as
$$
\eqalign{
       \biggl[ {\pi/2 \over \df^2 -\kappa}
           & \Bigl( \Pi^2_{\rho} -\df \Pi_{\df} \Pi_{\rho}
                  + {\kappa \over 4} \Pi^2_{\df} \Bigr)
           + {2 \over \pi}  \bigl(  \df \df^{\prime\prime}
                 -\df \df^{\prime} \rho^{\prime}
                 -\lambda^2 \df^2 \e^{2\rho} \bigr)                  \cr
           & -{\kappa \over 2\pi}
               \bigl( \rho^{\prime 2} -2\rho^{\prime\prime} \bigr)
           +\sum^N_{j=1}  \Bigl( \pi \Pi^2_{f_j}
                   + {1 \over 4\pi} f^{\prime 2}_j \Bigr)
                     \biggr]  \Psi =0                                 \cr
         }    \eqno(28)
$$
and
$$
      \bigl( \df^{\prime} \Pi_{\df} + \rho^{\prime} \Pi_{\rho}
              -\Pi^{\prime}_{\rho} +\sum^N_{j=1} \Pi_{f_j} f^{\prime}_j
                   \bigr) \Psi =0    ~,
       \eqno(29)
$$
where $\kappa $ is defined by eq.(20). $\Psi $ is a physical state.
The prime stands for the derivative with respect to the space coordinate.

  Here we have two remarks. The first is that the fields $\rho $ and $\df $
are dynamical variables so that it is significant to consider the equations
of motion of $\rho $ and $\df $. But $\bg $ is not dynamical. So we should
not regard the physical state conditions as the equations of motion of $\bg$.
The conditions come from the symmetry of the  theory. In this point
of view the conditions (28-29) indeed correspond to the \lq\lq constraints".

  The second remark is that the energy-momentum tensor
${\hat T}_{\alpha\beta} $ is transformed as non-tensor
because the Liouville
field $\rho $ is transformed as (23) for the conformal coordinate
transformation. In the light-cone
coordinate we get
$$
\eqalign{
         {\hat T}^{\pp}_{\pm\pm}(x^{\pp})
            & = \biggl( {\pd x^{\pm} \over \pd x^{\pm \pp} } \biggr)^2
              \bigl( {\hat T}_{\pm\pm}(x)
                    + {\kappa \over \pi} t_{\pm}(x) \bigr)  ~,   \cr
         {\hat T}^{\pp}_{+-}(x^{\pp})
            & = \biggl\vert {\pd x \over \pd x^{\pp}} \biggr\vert^2
              {\hat T}_{+-}(x)   ~,                                \cr
        }  \eqno(30)
$$
where $t_{\pm}(x)$ is the Schwarzian derivative
$$
     t_{\pm}(x) =  {\pd \gamma(x) \over \pd x^{\pm}}
                        {\pd \gamma(x) \over \pd x^{\pm}}
                         -{\pd^2 \gamma(x) \over \pd x^{\pm 2}} ~, \qquad
     \gamma(x)={1 \over 2} \log \biggl\vert
                          {\pd x^{\pp} \over \pd x}
                             \biggr\vert^2 ~.
   \eqno(31)
$$
Therefore the physical state conditions (28-29) correspond to the
case of $t_{\pm} =0 $. To determine what coordinate system corresponds to
this case is a physical requirement.
It is natural that the coordinate system which is asymptotically Minkowskian
is considered as the coordinate system with $t_{\pm} =0$.

    If we rewrite the canonical momentums as the differential operators
$$
     \Pi_{\rho} = {\delta \over i\delta \rho} ~, \quad
     \Pi_{\df} = {\delta \over i\delta \df}   ~, \quad
     \Pi_{f_j} = {\delta \over i\delta f_j}   ~,
        \eqno(32)
$$
the eqs.(28) and (29) give the differential equations similar to the
Wheeler-DeWitt  equations\footnote\dagger{See for example ref.10,
in which the spherically
symmetric gravitational system of 3+1 dimensions is
discussed. Application to the 1+1 dimensional dilaton gravity is
straightforward.}.
The most important difference between  the usual Wheeler-DeWitt equations and
ours is just the measures of the fields. In our case the commutation relations
are explicitly defined on the Minkowski background, but in ADM formalism
they are implicitly defined on the curved metric.
Therefore at first sight the conditions (28-29) seem to coincide with the
usual Wheeler-DeWitt equations at $\kappa=0$, but
this is wrong.

   If $ \kappa > 0 $, there is a singularity at finite $\df^2 =\kappa $.  The
region $\df^2 >\kappa $ is the classically allowed region\footnote\star{Here
${\hat I}$ is considered as a classical action}, whereas the region
$\kappa > \df^2 >0 $ is called the Liouville region, where
the sign of the kinetic term of eq.(28) changes.  This is the classically
forbidden region. The existence of the Liouville region is mysterious. There
may be some possibility of gravitational tunneling through this region.
If $\kappa <0 $, the situation drastically changes. In this
case the singularity  disappears.

{\bf 4. Black hole dynamics}

   Until now the arguments are completely non-perturbative. If we can solve
the physical state conditions exactly, the solution should include the
complete dynamics of black hole. Unfortunately it is a very difficult problem
so that we take an approximation. The original action (1) is order of
$1/\hbar $, but the Liouville part of ${\hat I}$ is zeroth order of $\hbar $.
However, if $\vert \kappa \vert $ is large enough, then it is meaningful to
consider the \lq\lq classical" dynamics of ${\hat I}$.  This is nothing but
the semi-classical approximation, which is valid only in the case of
$M \gg 1 $ and $N \gg 1$. In the other cases the quantum effect of gravitation
becomes important. The classical dynamics of ${\hat I}$ is ruled by the
equations ${\hat T}_{\alpha\beta} =0 $ and
the dilaton equation of motion
$$
\eqalign{
         -2 \pd_+ \df \pd_+ \df +2 \df \pd^2_+ \df
              -4 \df \pd_+ \df \pd_+ \rho
           & + {1 \over 2} \sum^N_{j=1} \pd_+ f_j \pd_+ f_j          \cr
            -\kappa &(\pd_+ \rho \pd_+ \rho -\pd^2_+ \rho +t_+ )=0 ~, \cr
         -2 \pd_- \df \pd_- \df +2 \df \pd^2_- \df
              -4 \df \pd_- \df \pd_- \rho
           & + {1 \over 2} \sum^N_{j=1} \pd_- f_j \pd_- f_j          \cr
            -\kappa & (\pd_- \rho \pd_- \rho -\pd^2_- \rho +t_- )=0 ~, \cr
         -2 \pd_+ \df \pd_- \df -2 \df \pd_+ \pd_- \df
               -\lambda^2 \df^2 \e^{2\rho} & = 0                      \cr
        }   \eqno(33)
$$
and
$$
        4 \pd_+ \pd_- \df +2\df \pd_+ \pd_- \rho
              +\lambda^2 \df \e^{2\rho} = 0  ~.
         \eqno(34)
$$
These are nothing but the CGHS equations\refmark{\cghs}
with the coefficient $\kappa$
instead of $N/12 $ before the Liouville part. Many authors have solved these
equations for $\kappa >0 $ and derived the dynamics of evaporating
black hole.\refmark{\rst,\w}
Initially the location of the horizon shifts to the out-side of the classical
horizon defined by the solution (3) by quantum effects
(almost matter's effects). Then the black hole evaporates and approaches to
the singularity\footnote\dagger{The location of the singularity given by
solving the equations (33) and (34) coincides with that determined by the
physical state conditions. Note that at the singularity the curvature is
singular, but the metric is not so.} asymptotically.
As far as the gauge-fixed action is treated
classically, the horizon does not seem to cross the singularity.
As mentioned before  the quantum mechanical region
$\kappa > \df^2 > 0 $ extends behind the singularity, where the quantum
gravitational effects become important.

   If $N $ is small, the \lq\lq non-anomalous" quantum corrections of gravity
part maybe cannot neglect and the approximation becomes bad. Nevertheless we
apply the approximation for $\kappa < 0 $ because we hope that some new
insights are obtained from the solution. If $\kappa < 0 $, the singularity
disappears. The location  of the horizon initially shifts to the inside of
the classical horizon. If the effective mass of the black hole  is defined
by $M_{BH} = \lambda \df^2 \vert_{\hbox{horizon}}$, this means that the
initial mass of the black hole is less than the infalling matter flux $M $.
After the black hole is formed, the positive flux comes in through the horizon
and the black hole mass increases. It seems that the horizon approaches to the
classical horizon asymptotically.

   The problem of the information loss seems to come out in the case of
$\kappa > 0 $.
Then the black hole evaporates and the information seems to be lost.
However in this case the Liouville region extends behind the singularity.
So it appears that  there is a possibility that
the informations run away through this region by gravitational tunneling.
On the other hand, if $\kappa \leq 0 $, the Liouville region disappears.
But the black hole seems to be stable. In this case the problem of the
information loss appears not to  exist.

\ack{The author would like to thank Asato Tsuchiya and Tamiaki Yoneya for
valuable
discussion. This work is supported in part by Soryuushi Shogakukai.}

\refout

\bye